\RequirePackage{fix-cm}
\RequirePackage{amssymb}
\documentclass[twocolumn,epjc3]{svjour3}  
\smartqed  
\RequirePackage{graphicx}
\journalname{Fortschritte der Physik}
\begin{document}

\title{Cosmology from non-minimal geometry-matter coupling}

\author{{B.S. Gon\c{c}alves}\thanksref{e1}$^{,1}$
        \and
        {P.H.R.S. Moraes}\thanksref{e2}$^{,2}$
        \and
        {B. Mishra}\thanksref{e3}$^{,3}$}

\thankstext{e1}{e-mail: goncalvesbs.88@gmail.com}
\thankstext{e2}{e-mail: moraes.phrs@gmail.com}
\thankstext{e3}{e-mail: bivu@hyderabad.bits-pilani.ac.in}

\institute{{Instituto Tecnol\'ogico de Aeron\'autica - Departamento de F\'isica, 12228-900, S\~ao Jos\'e dos Campos, SP, Brazil \\$^2$Universidade Federal do ABC (UFABC) - Centro de Ci\^encias Naturais e Humanas (CCNH) - Avenida dos Estados 5001, 09210-580, Santo Andr\'e, SP, Brazil} \\$^3$Department of Mathematics,
Birla Institute of Technology and Science-Pilani, Hyderabad Campus,
Hyderabad-500078, India}

\date{Received: date / Accepted: date}

\maketitle

\begin{abstract}

We construct a cosmological model from the inception of the Friedmann-Lem\^aitre-Robertson-Walker metric into the field equations of the $f(R,L_m)$ gravity theory, with $R$ being the Ricci scalar and $L_m$ being the matter lagrangian density. The formalism is developed for a particular $f(R,L_m)$ function, namely $R/16\pi +(1+\sigma R)L_{m}$, with $\sigma$ being a constant that carries the geometry-matter coupling. Our solutions are remarkably capable of evading the Big-Bang singularity as well as predict the cosmic acceleration with no need for the cosmological constant, but simply as a consequence of the geometry-matter coupling terms in the Friedmann-like equations. 

\end{abstract}

\section{Introduction}\label{sec:int}

According to the Planck Satellite, $\sim95\%$ of the universe is made by dark energy and dark matter, and only the remaining $\sim5\%$ is made by well known baryonic matter \cite{planck_collaboration/2016}.  

Dark energy is the name given for what causes the universe expansion to accelerate. We know for more than $90$ years that the universe is expanding \cite{hubble/1929}. However, by the end of the last century, observations of supernova Ia brightness diminishing indicated that such an expansion occurs in an accelerated way \cite{riess/1998,perlmutter/1999}. Such a dynamical feature is highly counter-intuitive since one expects gravity, as an attractive force, to slow down the expansion of the universe.

In the standard cosmological model, the cosmic acceleration is said to be caused by the presence of the cosmological constant $\Lambda$ in the Einstein's field equations of General Relativity (GR), 

\begin{equation}\label{i1}
G_{\mu\nu}=8\pi T_{\mu\nu}-\Lambda g_{\mu\nu}.
\end{equation}
In (\ref{i1}), $G_{\mu\nu}$ is the Einstein tensor, $T_{\mu\nu}$ is the energy-momentum tensor, $g_{\mu\nu}$ is the metric and units such that $G_N=c=1$ are assumed throughout the article, with $G_N$ being the Newtonian gravitational constant and $c$ the speed of light.

The cosmological constant $\Lambda$ is physically interpreted as the vacuum quantum energy \cite{weinberg/1989}, which has the repulsive character needed for accelerating the expansion. This vacuum quantum energy is expected to make up $\sim70\%$ of the universe \cite{planck_collaboration/2016}.

The remaining $\sim25\%$ of the dark sector pointed by the Planck Satellite is expected to be made by dark matter \cite{planck_collaboration/2016}. Dark matter is a kind of matter that does not interact electromagnetically and therefore cannot be seen. However, it has a key gravitational role, such as in the large-scale structures formation in the universe \cite{padmanabhan/1993} and in the behavior of spiral  galactic rotation curves \cite{kent/1986,kent/1987}.

The above scenario is indeed the best one to fit observational data. However, it is ``haunted'' by the cosmological constant problem \cite{weinberg/1989} and the non-detection of dark matter particles \cite{baudis/2016}. While the dark matter non-detection problem may eventually be solved by increasing the experiments precision and technological capacity, the cosmological constant problem is quite serious and sometimes is referred to as {\it the worst prediction of Theoretical Physics}. The cosmological constant problem is the fact that the value of the cosmological constant obtained from Particle Physics methods  \cite{weinberg/1989} is $\sim120$ orders of magnitude different from the value needed to fit cosmological observations \cite{planck_collaboration/2016,riess/1998,perlmutter/1999}. 

This has led a great number of theoretical physicists to search for alternatives to describe the cosmic acceleration with no need for invoking the cosmological constant. Those alternatives are the Extended Gravity Theories (EGTs) (also referred to as alternative or modified gravity theories).
 
EGTs {\it extend} GR to incorporate new degrees of freedom. They normally substitute the Ricci scalar $R$ in the Einstein-Hilbert gravitational action

\begin{equation}\label{i2}
	S=\frac{1}{2}\int d^4x\sqrt{-g}R,
\end{equation}
with $g$ being the metric determinant, by a general function of $R$ and/or other scalars. For a review on the subject of EGTs, one can check \cite{capozziello/2011}.

For instance, the $f(R)$ theories of gravity \cite{sotiriou/2010} substitute $R$ by a generic function $f(R)$ in (\ref{i2}). When applying the variational principle in the resulting action, the new terms appearing in the field equations can account for dark energy and even dark matter effects \cite{capozziello/2008}.


Here we choose to work with a sort of expansion of the $f(R)$ gravity, namely $f(R,L_{m})$ gravity \cite{harko/2010}, with $L_{m}$ being the matter lagrangian density, which is motivated by some $f(R)$ theory shortcomings (check \cite{joras/2011}, for instance). This theory allows, through the substitution of $R$ by $f(R,L_{m})$ in (\ref{i2}), to generalize not only the geometrical but also the material sector of a theory. It also allows the possibility of non-minimally coupling geometry and matter in a gravity theory, say, from a product of $R$ and $L_{m}$ in $f(R,L_{m})$.

There are some important applications of theories with non-minimal geometry-matter coupling (GMC) to be seen in the literature nowadays. In \cite{banados/2010}, Ba\~nados and Ferreira discovered that it is possible to avoid the big-bang singularity through GMC. The dark energy and dark matter issues were approached in GMC theories in \cite{moraes/2017} and \cite{harko/2010b}, respectively. For a review about GMC gravity, we suggest \cite{harko/2014}.

Particularly, the $f(R,L_{m})$ gravity, to be approached here, also contains some important applications. We remark a few of them below. A dynamical system approach was used in \cite{azevedo/2016} to analyze the viability of some $f(R,L_{m})$ dark energy candidates. The energy conditions were applied to this theory in \cite{wang/2012} and wormholes solutions were obtained in \cite{garcia/2011,garcia/2010}.

Our intention in the present article is to construct a homogeneous and isotropic cosmological model in the $f(R,L_{m})$ gravity. In order to do so, we will choose to work with the functional form $f(R,L_{m})=R/16\pi +(1+\sigma R)L_{m}$, as it was done in \cite{garcia/2011,garcia/2010}. In such a functional form, $\sigma$ is a parameter that controls the coupling between geometry and matter. When $\sigma=0$ one automatically recovers GR, as it will be shown in the next section.

\section{The $f(R,L_{m})$ gravity}\label{sec:frl}

\hspace{4.mm} Working with the $f(R,L_{m})$ formalism, we start from the action \cite{harko/2010}

\begin{equation}\label{mgcg1}
S=\int d^ 4x\sqrt{-g}f(R,L_{m}).
\end{equation}
By varying the above action with respect to the metric $g_{\mu\nu}$ yields the following field equations:

\begin{eqnarray}\label{1}
f_{_{R}}R_{\mu\nu} + \left(g_{\mu\nu}\nabla_{\mu}\nabla^{\mu} -\nabla_{\mu}\nabla_{\nu}\right)f_{_{R}}-\frac{1}{2}\big(\!&f&-f_{_{L_{m}}}L_{m}\big)g_{\mu\nu}\nonumber \\
&=&\frac{1}{2}f_{_{L_{m}}}T_{\mu\nu},
\end{eqnarray}

\vspace{1.mm}
\noindent with $f_{_{R}}\equiv\partial f(R,L_{m})/\partial R$,\hspace{1.mm} $f_{_{L_{m}}}\equiv\partial f(R,L_{m})/\partial L_{m}$ and $f=f(R,L_{m})$. Moreover, the energy-momentum tensor reads as

\vspace{1.mm}

\begin{eqnarray}\label{9}
T_{\mu\nu}= -\frac{2}{\sqrt{-g}}\frac{\delta \left(\sqrt{-g}L_{m}\right)}{\delta g^{\mu\nu}}=g_{\mu\nu}L_{m} -2\frac{\partial L_{m}}{\partial g^{\mu\nu}},
\end{eqnarray}
assuming that the lagrangian density of matter depends only on the components of the metric tensor, and not on its derivatives.

\vspace{3.mm}
Taking the covariant derivative in Eq.(\ref{1}), we obtain
\begin{eqnarray}\label{2}
\nabla^{\mu}T_{\mu\nu}= 2\left(\nabla^{\mu}\ln f_{_{L_{m}}}\right)\frac{\partial L_{m}}{\partial g^{\mu\nu}}.
\end{eqnarray}

\subsection{The $f(R,L_{m})=R/16\pi+(1+\sigma R)L_{m}$ gravity}\label{ss:frl}

\hspace{4.mm} Assuming $f(R,L)=R/16\pi +\left(1+\sigma R\right)L_{m}$ yields, for the field equations (\ref{1}), the following
\begin{eqnarray}\label{3}
G_{\mu\nu}=\hat{T}_{\mu\nu},
\end{eqnarray}
where $G_{\mu\nu}$ is the Einstein tensor and $\hat{T}_{\mu\nu}$ is defined as 
\begin{eqnarray}
\hat{T}_{\mu\nu}=8\pi T_{\mu\nu}-16\pi\!&\sigma&\!\Bigg[L_{m}R_{\mu\nu}-\frac{1}{2}R\, T_{\mu\nu}\\ \nonumber
&+&\left(g_{\mu\nu}\nabla_{\mu}\nabla^{\mu} -\nabla_{\mu}\nabla_{\nu}\right)L_{m} \Bigg],
\end{eqnarray}
with $R_{\mu\nu}$ representing the Ricci tensor.

From Eq.(\ref{2}), we obtain, as the covariant derivative of the energy-momentum tensor, the following

\begin{eqnarray}\label{8}
\nabla^{\mu}T_{\mu\nu}=2\left(\frac{\sigma}{1+\sigma R}\right)\nabla^{\mu}R\, \frac{\partial L_{m}}{\partial g^{\mu\nu}}.
\end{eqnarray}

\section{Cosmology from non-minimal geometry-matter coupling}\label{sec:cgmc}

\subsection{Friedmann-like equations}\label{ss:fle}

\hspace{4.mm} To construct the Friedmann-type equations, we assume the Friedmann-Lem\^aitre-Robertson-Walker flat metric, which agrees with the recent observations of the Planck satellite \cite{planck_collaboration/2016}, 
\begin{equation}\label{fle1}
ds^ 2=dt^ 2-a(t)^ 2[dr^ 2+r^ 2(d\theta^ 2+\sin^ 2\theta d\phi^ 2)],
\end{equation}
and the energy-moment tensor of a perfect fluid,
\begin{equation}\label{fle2}
T_{\mu\nu}=\left(\rho +p\right)u_{\mu}u_{\nu}-pg_{\mu\nu}
\end{equation}
with $a(t)$ being the scale factor, which dictates how distances evolve in the universe and $u_{\mu}$, $\rho$ and $p$, respectively, represent the four-velocity, the density of matter-energy and the pressure of the fluid that describes the universe content.

\vspace{3.mm}
Developing the field equations (7)-(8) for (10)-(11), we obtain

\begin{eqnarray}\label{4}
3\!\left(\frac{\dot{a}}{a}\right)^{2}\!=\!8\pi\rho\!-\!48\pi \sigma\Bigg[\dot{L_{m}}\frac{\dot{a}}{a}\!-\! L_{m}\frac{\ddot{a}}{a}\!+\!\left(\frac{\ddot{a}}{a}+\frac{\dot{a}^{2}}{a^{2}}\right)\!\rho\Bigg],
\end{eqnarray}

\begin{eqnarray}\label{5}
2\frac{\ddot{a}}{a}+\left(\frac{\dot{a}}{a}\right)^{2}=-8\pi p +16\pi\sigma\,\Bigg[L_{m} \left(\frac{\dot{a}}{a}\right)^{2}-\ddot{L_{m}}\nonumber\\
-2\dot{L_{m}} \frac{\dot{a}}{a}+\left(L_{m}+3p \right)\left(\frac{\ddot{a}}{a}+\frac{\dot{a}^{2}}{a^{2}}\right)\Bigg],
\end{eqnarray}
with dots representing time derivatives.

\subsection{Cosmological solutions}\label{ss:cs}

\hspace{4.mm} In view of a characteristic description for the material lagrangian of the present model, some discussions can be seen \cite{far/2009,berto/2008,hm/2020}. It seems there is a natural direction to take $L_{m}\!=\!-p$ \footnote{It is worth mentioning that variation in the sign of $p$ is generally associated with a respective change in the metric signature (with some exceptions).} \cite{carv/2017,velt/2017,schutz/1970}, which we are going to follow here.


We will also assume a dust-like matter component to dominate the universe dynamics. By doing so we do not need an exotic fluid, such as dark energy, with $p/\rho=-1$, in order to describe the present acceleration of the universe expansion. If we are anyhow able to predict such a dynamical feature, it comes purely from the GMC terms of the theory. The Friedmann-like equations now read

\begin{eqnarray}\label{6}
\left(\frac{\dot{a}}{a}\right)^{2}= \frac{8\pi}{3}\rho -16\pi\sigma \left(\frac{\ddot{a}}{a}+\frac{\dot{a}^{2}}{a^{2}}\right)\rho,
\end{eqnarray}

\begin{eqnarray}\label{7}
\frac{\ddot{a}}{a}=-\frac{4\pi}{3}\rho + 8\pi \sigma \left(\frac{\ddot{a}}{a}+\frac{\dot{a}^{2}}{a^{2}}\right)\rho.
\end{eqnarray}
It is worth noting, as mentioned in Eqs.(\ref{3})-(8), that by making $\sigma=0$ in (\ref{6}) and (\ref{7}), the usual Friedmann equations that describe the dynamics of a matter-dominated universe are recovered.

\vspace{2.mm}
By manipulating the above equations, we obtain the following differential equation for the scale factor
\begin{eqnarray}\label{11}
\frac{\ddot{a}}{a} + \left(\frac{\dot{a}}{a}\right)^{2}=\Sigma,
\end{eqnarray}
where $\Sigma=(4\pi\rho/3)/\left(1+8\pi\sigma\rho\right)$.
\vspace{5.mm}

Moreover, Eq.(\ref{8}) now reads
\begin{eqnarray}\label{10}
\dot{\rho}+3\frac{\dot{a}}{a}\rho=0,
\end{eqnarray}
which indicates conservation of matter-energy in the present formalism.

Now, one can find 
\begin{eqnarray}\label{12}
\rho(t)=\varphi \left(\frac{1}{t+\mathcal{C}}\right)^{2},
\end{eqnarray}
as a solution for $\rho$, with $\varphi$ and $\mathcal{C}$ being constants. 

\vspace{3.mm}
Under the condition $a(0)=0$ and assuming
$\mathcal{C}=0$ for simplicity, in possession of the result (\ref{12}), we can solve the scale factor equation (\ref{11}), yielding
\begin{eqnarray}\label{13}
a(t)= \zeta\:\sqrt{F[\alpha] \left( \varphi\,\sigma\right)^{-1/2}\: t},
\end{eqnarray}

\noindent with $\zeta$ a constant, $\alpha=1/4$,\hspace{1.mm}$F[n]={}_2 F_{1}\Bigg[n-\delta,n\\ +\delta,\displaystyle\left(n+\frac{5}{4}\right), -\frac{t^{2}}{8\pi \,\varphi\sigma} \Bigg]$, $\delta= \sqrt{1/16+2\pi\varphi/3}$ and 

${}_2F_{1}\left[a,b,c,d\right]=\sum_{n=0}^{\infty}\frac{\mathbf{(a)_{n}}\,\mathbf{(b)_{n}}}{\mathbf{(d)_{n}}}\frac{(c)^{n}}{n!}$ is a hypergeometric function\footnote{Known as  Gaussian hypergeometric series.}, where the terms in bold evolve according to the Pochhammer Symbol \cite{sea/1991}.

Fig.\ref{fig:a1t} below shows the temporal behavior developed by the scale factor by modifying the free parameters $\varphi$ and $\sigma$.


\begin{figure}[h!]
	\centering
	\includegraphics[width=1.02\linewidth]{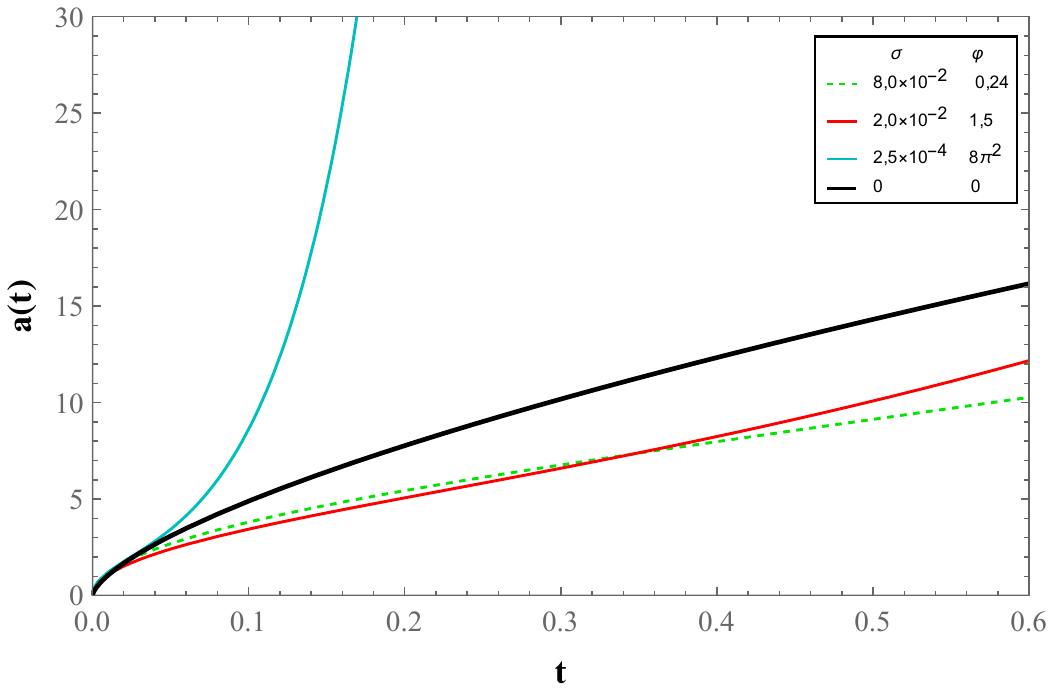}
	\caption{Evolution of the scale factor as a function of time, with $\zeta=7\pi/5$,  under the analysis of different $\varphi$ and $\sigma$ values.}
	\label{fig:a1t}
\end{figure}

\begin{figure}[h!]
	\centering
	\includegraphics[width=1.02\linewidth]{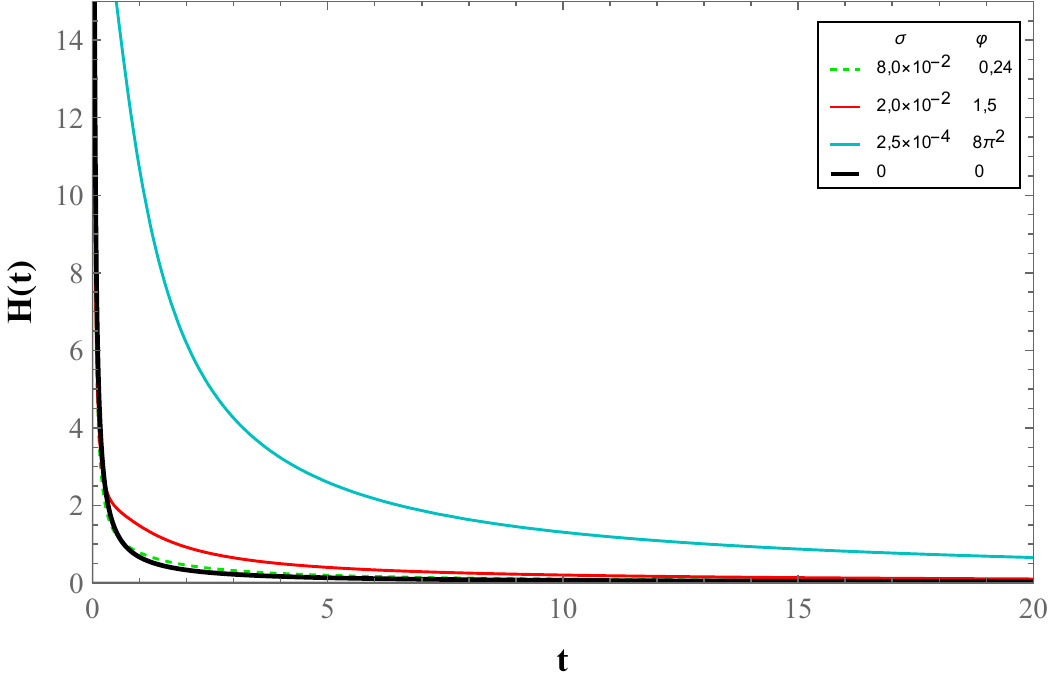}
	\caption{Time evolution of the Hubble parameter, with $\zeta=7\pi/5$, under the analysis of different $\varphi$ and $\sigma$ values.}
	\label{fig:h1t}
\end{figure}

After finding the solution for the scale factor, we obtain the solution for the Hubble parameter $H=\dot{a}/a$ as
\begin{eqnarray}\label{14}
H(t)=\frac{1}{2t} +\frac{t}{18\, \sigma}\left(\frac{F[\beta]}{F[\alpha]}\right),
\end{eqnarray}
with $\beta=5/4$. The evolution of the Hubble parameter in time can be seen in Figure \ref{fig:h1t}.

The deceleration parameter $q(t)=-\ddot{a}a/\dot{a}^{2}$ is responsible for classifying whether the universe is in the process of acceleration, $q<0$, or deceleration , $q>0$. In our model it reads as

\begin{eqnarray}\label{15}
q(t)\!=\!\Bigl\{\!\left(9\sigma F[\alpha]\right)^{2}\!-\!(6t)^{2}\sigma F[\alpha] F[\beta]\!+\!\left(t^{2}F[\beta]\right)^{2}
+3t^{4}\nonumber\\
\left(9\!-\!4\pi \varphi\right)\!F[\alpha] F[\gamma]/10\pi\varphi\Bigr\}\!\Big{/}\!\left(9\sigma F[\alpha]\!+\!t^{2}\!F[\beta]\right)^{2},
\end{eqnarray}
with $\gamma=9/4$. 

\begin{figure}[h!]
	\centering
	\includegraphics[width=1.02\linewidth]{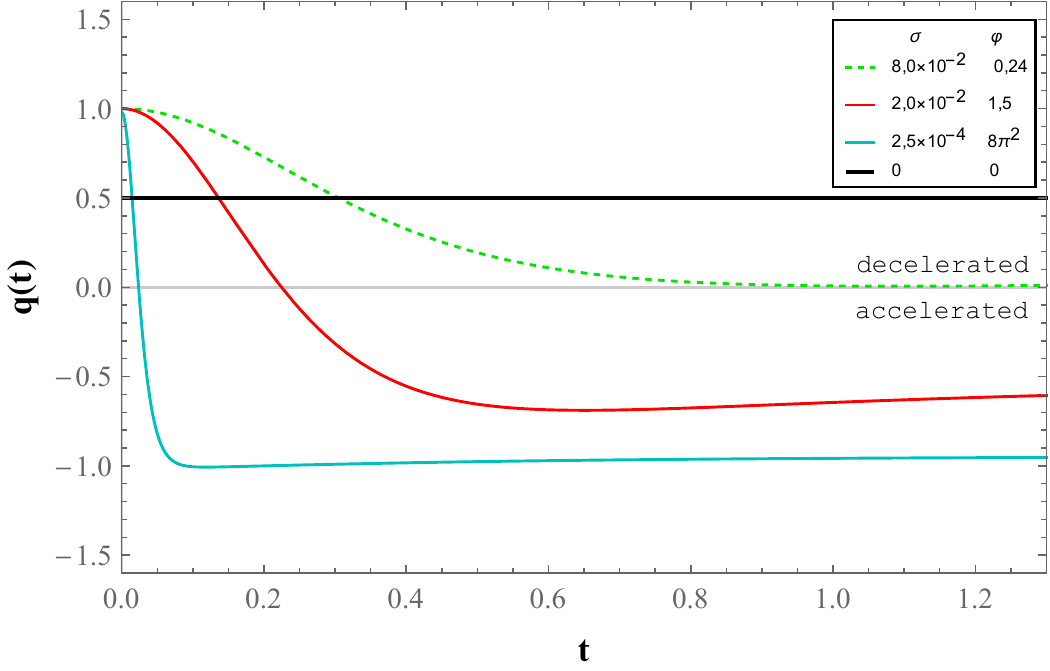}
	\caption{Time evolution of the deceleration parameter, with $\zeta=7\pi/5$, under the analysis of different $\varphi$ and $\sigma$ values.}
	\label{fig:q1t}
\end{figure}

From the equation above, we can see the temporal evolution of the deceleration parameter in Figure \ref{fig:q1t}.

Next we will interpret the solutions obtained for the cosmological parameters.

\subsection{Cosmological interpretations}\label{ss:ci}

First of all, it is interesting to mention that solution (18) for $\rho$ is elegantly capable of evading the Big-Bang singularity. Note that for $\mathcal{C}\neq0$, $\rho$, in principle, does not diverge for $t=0$. It is well-known that this is not the case for standard cosmology \cite{barb/2003}. This interesting result recovers what was recently  shown by Ba\~nados and Ferreira in \cite{banados/2010}, that is, GMC models are capable of elegantly evade the Big-Bang singularity. 

Now let us analyze the scale factor $a(t)$ and the Hubble parameter $H(t)$, which respectively read according to Eqs.(19) and (20) and whose time evolutions appear in Figs.1 and 2. From Fig.1 we can see that the scale factor as well as its time derivatives are always positive. This is in agreement with an expanding universe and can also be observed in the Fig.2 features. The latter shows that the Hubble parameter is positive and decreasing with time. Those features are in agreement with what is predicted by observations and with the standard cosmology model \cite{planck_collaboration/2016,barb/2003}. Moreover note that as time passes by, $H$ tends to a constant. From the Hubble parameter definition, a constant $H$ is an indication of an exponential scale factor, which is required for explaining the cosmic acceleration.

 Finally, Figure 3 shows the evolution of the deceleration parameter for different values of the free parameters of the model. Particularly, the black solid line recovers exactly what is expected in a matter-dominated universe governed by GR. In other words, $q=0.5$ is exactly what one obtains when solving the standard Friedmann equations for $p=0$. 
 
 The values assumed for the free parameters $\varphi$ and $\sigma$ in the green curve do not allow the universe to transit from a decelerated to an accelerated regime of expansion
 
 Otherwise, we have also the light blue and red curves, which clearly indicate a transition from a decelerated ($q>0$) to an accelerated ($q<0$) stage of the universe expansion. As the green curve, both departure from $1$, which indicate a primordial radiation-dominated universe, but naturally assume negative values as time passes by, indicating that in the present model the universe expansion accelerates simply as a consequence of the GMC model features. Moreover, note that the values assumed for the free parameters in the light blue curve yield a de Sitter-like universe in the future. Note also that both light blue and red curves eventually assume the present value estimated for $q$, which according to \cite{giostri/2012} is $q_0=-0.31\pm0.11$, while according to \cite{lu/2008} is $q_0=-0.64\pm0.15$.

\section{Final remarks}\label{sec:fr}

We have obtained a matter-dominated GMC model of cosmology from the $f(R,L_m)$ gravity formalism. The Friedmann-like equations were obtained for the model, as well as the continuity equation. Then we have obtained analytical solutions for all the cosmological parameters. 

Remarkably, our solution for $\rho$ evades the Big-Bang singularity, which now seems to be a profitable feature of GMC models (check \cite{banados/2010}).  Not only the present model was able to evade the Big-Bang singularity, it was also capable of describing the transition from a decelerated to an accelerated stage of the universe expansion. This can be clearly seen in Fig.3, that shows two possibilities for the deceleration parameter to assume negative values, in accordance with observational estimates.

The cosmic acceleration is one among many challenges theoretical physicists face nowadays. The analysis of the rotation curves of galaxies is a natural next step to test the present formalism. That is, in the same way the present GMC model is capable of describing the dark energy effects, could it also be capable of describing the dark matter effects in the galactic scales? We shall investigate and report that soon.

It should also be stressed here that the $f(R,L_m)$ theory has already shown good results in the physics of wormholes. While GR wormholes need to be filled by exotic negative-mass matter, the $f(R,L_m)$ wormholes do not \cite{garcia/2011,garcia/2010}. Last, but definitely not least, the very same model used here to describe the dynamics of the universe was used in \cite{carvalho/2020} to obtain hydrostatic equilibrium configurations of neutron stars. The theory remarkably makes possible to describe pulsars as massive as PSR J2215+5135 \cite{linares/2018} from a simple equation of state for nuclear matter.

\bigskip

\begin{acknowledgements}
BSG would like to thank CAPES for financial support. BM thanks IUCAA, Pune, India for providing academic support through visiting associateship program. The authors are thankful to the honourable referees for the comments and suggestions for the improvement of the paper.

\end{acknowledgements}

\end{document}